\pdfoutput=1

\documentclass[11pt,twoside,a4paper,cmspaper,final,collab]{cms-tdr}
\def\svnVersion{240606}\def\cmsCernNo{2013-211}\def\cmsCernDate{\today}\def\cmsMessage{Published in Physical Review Letters as \href{http://dx.doi.org/10.1103/PhysRevLett.112.182001}{\doi{10.1103/PhysRevLett.112.182001.}}}

\begin{document}\cmsNoteHeader{TOP-13-003}

\hyphenation{had-ron-i-za-tion}
\hyphenation{cal-or-i-me-ter}
\hyphenation{de-vices}

\RCS$Revision: 231684 $
\RCS$HeadURL: svn+ssh://svn.cern.ch/reps/tdr2/papers/TOP-13-003/trunk/TOP-13-003.tex $
\RCS$Id: TOP-13-003.tex 231684 2014-03-14 18:43:18Z linacre $
\cmsNoteHeader{TOP-12-004} % This is over-written in the CMS environment: useful as preprint no. for export versions
\title{\texorpdfstring{Measurements of \ttbar Spin Correlations and Top-Quark Polarization Using Dilepton Final States in \pp Collisions at $\sqrt{s}=7\TeV$}{Measurements of t t-bar Spin Correlations and Top-Quark Polarization Using Dilepton Final States in pp Collisions at sqrt(s) = 7 TeV}}

\providecommand{\ee}{\ensuremath{\Pe\Pe}\xspace}
\newcommand{\emu}{\ensuremath{\Pe\Pgm}\xspace}
\newcommand{\mumu}{\ensuremath{\Pgm\Pgm}\xspace}

\newcommand{\eepm}{\ensuremath{\Pep\Pem}\xspace}
\newcommand{\mmpm}{\ensuremath{\Pgmp \Pgmm}\xspace}
\newcommand{\ttpm}{\ensuremath{\Pgt^+ \Pgt^-}\xspace}
\newcommand{\empm}{\ensuremath{\Pe^\pm \Pgm^\mp}\xspace}
\newcommand{\pp}{\ensuremath{\Pp\Pp}\xspace}
\newcommand{\ppbar}{\ensuremath{\Pp\Pap}\xspace}
\newcommand{\ase}[2]{\ensuremath{_{~- #1}^{~+ #2}}}
\newcommand{\roots}{\ensuremath{\sqrt{s}}}
\newcommand{\lhcE}[1]{\ensuremath{\roots ={#1}\TeV}}
\newcommand{\dy}{\ensuremath{\cPZ/\Pgg^\star}+jets}
\newcommand{\dyee}{\ensuremath{\dy\to\eepm}}
\newcommand{\dymm}{\ensuremath{\dy\to\mmpm}}
\newcommand{\dytt}{\ensuremath{\dy\to\ttpm}}
\newcommand{\mll}{\ensuremath{M_{\ell\ell}}}
\newcommand{\isocomb}{\ensuremath{I_\text{comb}}}
\newcommand{\WoZ}{\ensuremath{\PW/\PZ}}
\providecommand{\POWHEG} {\textsc{Powheg}\xspace}
\providecommand{\PYTHIA} {\textsc{Pythia}\xspace}
\providecommand{\HERWIG} {\textsc{Herwig}\xspace}
\newcommand{\tW}{\ensuremath{\cPqt\PW}\xspace}
\newcommand{\VV}{\ensuremath{\cmsSymbolFace{VV}}}
\newcommand{\wmn}{\ensuremath{\PW\to\Pgm\Pgngm}\xspace}

\newcommand{\ptll} {\ensuremath{\pt(\ell\ell)}}
\newcommand{\tmass}{\ensuremath{M_{\cPqt}}\xspace}
\newcommand{\ttbarmass}{$m_{\ttbar}$}
\newcommand{\ttbareta}{$\abs{y_{\cPqt}+y_{\cPaqt}}$}
\newcommand{\ttll}{\ensuremath{\ttbar\to\ell^+\ell^-}\xspace}
\newcommand{\tttau}{\ensuremath{\ttbar\to\ell^{\pm}\Pgt^{\mp}/\Pgt^+\Pgt^-}\xspace}
\newcommand{\ttfake}{\ensuremath{\ttbar\to\text{other}}\xspace}
\newcommand{\wjets}{\ensuremath{\PW+\text{jets}}\xspace}

\ifthenelse{\boolean{cms@external}}
{\providecommand{\suppMaterial}{the Supplemental Material
[URL will be inserted by publisher]\xspace}}
{\providecommand{\suppMaterial}{Appendix~\ref{app:suppMat}\xspace}}

\date{\today}

\abstract{Spin correlations and polarization in the top quark--antiquark system are measured using dilepton final states produced in \pp collisions at the LHC at $\sqrt{s}=7\TeV$.
The data correspond to an integrated luminosity of 5.0\fbinv collected with the CMS detector.
The measurements are performed using events with two oppositely charged leptons (electrons or muons), a significant imbalance in transverse momentum, and two or more jets, where at least one of the jets is identified as originating from a \cPqb~quark.
The spin correlations and polarization are measured through asymmetries in angular distributions of the two selected leptons, unfolded to the parton level.
All measurements are found to be in agreement with predictions of the standard model.}

\hypersetup{%
pdfauthor={CMS Collaboration},%
pdftitle={Measurements of t t-bar Spin Correlations and Top-Quark Polarization Using Dilepton Final States in pp Collisions at sqrt(s) = 7 TeV},%
pdfsubject={CMS},%
pdfkeywords={CMS, physics, top, dilepton, spin correlation, polarization, asymmetry}}

\maketitle %maketitle comes after all the front information has been supplied

Measurements of spin correlations and polarization in the top quark--antiquark (\ttbar) system
provide direct access to the properties of the
bare top quark, as well as a test of the validity of perturbative quantum chromodynamics (QCD) in the \ttbar\ production process~\cite{ref:topquarkphysics}.
Such measurements are of particular interest given the anomalies in the \ttbar\ forward-backward production asymmetry observed at the Tevatron~\cite{ref:D0afb,ref:CDFafb2013}.
The top-quark lifetime has been measured as $3.29^{+0.90}_{-0.63} \times 10^{-25}\unit{s}$~\cite{PhysRevD.85.091104},
much shorter than the spin decorrelation time scale of $m_\cPqt/\Lambda^{2}_{\text{QCD}}\approx 10^{-21}\unit{s}$~\cite{MahlonParke2010},
where $m_\cPqt$ is the top-quark mass, measured as $173.20 \pm 0.87\GeV$~\cite{Aaltonen:2012ra}, and $\Lambda_{\text{QCD}}$ is the QCD scale parameter.
Consequently, the information about the spin of the top quark at production
is transferred directly to its decay products and can be accessed from their angular distributions.
In the standard model (SM),
top quarks are produced with a small amount of polarization arising from electroweak corrections to the QCD-dominated production process.
For models beyond the SM, couplings of the top quark to new particles can alter both the polarization of the top quark and the amount of spin correlation in the \ttbar\ system~\cite{ref:Krohn,ref:Fajfer,1742-6596-447-1-012015}.

At the Large Hadron Collider (LHC), top quarks are produced abundantly, mainly in pairs.
For low \ttbar\ invariant masses, the production
is dominated by the fusion of pairs of gluons with the same helicities, resulting in the creation of top-quark pairs with antiparallel spins.
At larger invariant masses, the dominant production is via the fusion of gluons with opposite helicities, resulting in \ttbar\ pairs with parallel spins.
These have the same configuration as \ttbar\ events produced
via $\cPq\cPaq$ annihilation~\cite{MahlonParke2010}.

In the decay $\ttbar \to  \ell^{+} \cPgn \cPqb  \ell^{-} \cPagn \cPaqb$, in the laboratory frame, the
difference in azimuthal angles of the charged leptons ($\Delta\phi_{\ell^+\ell^-}$) is sensitive to \ttbar\ spin correlations
and can be measured precisely without reconstructing the full event kinematics~\cite{MahlonParke2010}.
The top-quark spin can also be studied using $\theta^{\star}_\ell$, which is the angle of a charged lepton in the rest frame of its parent top quark or antiquark, measured in the helicity frame (i.e., relative to the direction of the parent quark in the \ttbar\ center-of-momentum frame).
The CDF, D0, and ATLAS spin correlation and polarization measurements used template fits to angular distributions and observed results consistent with SM expectations~\cite{CDFSpinCor,D0SpinCor,D0SpinCorMEM,AtlasSpinCor2012,PhysRevD.87.011103,Aad:1564320}.
In this analysis, the measurements are made using angular asymmetry variables unfolded to the parton level, allowing direct comparisons between the data and theoretical predictions.

The top-quark polarization $P$ in the helicity basis is given by $P = 2A_{P}$, where the asymmetry variable $A_{P} $ is defined as
\begin{equation*}
A_{P} = \frac{N \left[\cos(\theta^{\star}_\ell) > 0\right]-N \left[\cos(\theta^{\star}_\ell) < 0\right]}{N \left[\cos(\theta^{\star}_\ell) > 0\right]+N \left[\cos(\theta^{\star}_\ell) < 0\right]}.
\end{equation*}
Here the number of events $N$ is counted using the $\theta^{\star}_\ell$ measurements of both positively and negatively charged leptons ($\theta^{\star}_{\ell^+}$ and $\theta^{\star}_{\ell^-}$), assuming CP invariance.

For \ttbar\ spin correlations, the variable
\begin{equation*}
A_{\Delta\phi} = \frac{N (  \Delta \phi_{\ell^+\ell^-} > \pi/2 )-N ( \Delta \phi_{\ell^+\ell^-} < \pi/2 )}{N ( \Delta  \phi_{\ell^+\ell^-} > \pi/2 ) +N ( \Delta \phi_{\ell^+\ell^-} < \pi/2 )}
\end{equation*}
provides excellent discrimination between correlated and uncorrelated \cPqt\ and \cPaqt\ spins, while the variable
\begin{equation*}
A_{c_{1}c_{2}} = \frac{N (c_{1} c_{2} > 0)-N (c_{1} c_{2} < 0)}{N (c_{1} c_{2} > 0)+N (c_{1} c_{2} < 0)},
\end{equation*}
where $c_{1} = \cos(\theta^{\star}_{\ell^+}\!)$ and $c_{2} = \cos(\theta^{\star}_{\ell^-}\!)$,
provides a direct measure of the spin correlation coefficient $C_{\mathrm{hel}}$ using the helicity angles of the two leptons in each event: $C_{\mathrm{hel}} = -4 A_{c_{1}c_{2}}$~\cite{Bernreuther2013115}.

The results presented in this Letter are based on data that correspond to an integrated luminosity of 5.0\fbinv of proton-proton (\pp) collisions at $\sqrt{s}=7\TeV$, provided by the LHC and recorded by the Compact Muon Solenoid (CMS) detector in 2011.

The central feature of the CMS apparatus is a superconducting solenoid,
13\unit{m} in length and 6\unit{m} in diameter, which provides an axial magnetic
field of 3.8\unit{T}.  The bore of the solenoid is equipped with a variety of
particle detection systems. Charged-particle trajectories are
measured with silicon pixel and strip trackers covering the pseudorapidity region
$\abs{\eta}<2.5$, where $\eta =-\ln[\tan{\theta/2}]$, with $\theta$ the
polar angle of the trajectory of the particle with respect to the counterclockwise-beam direction.
A crystal electromagnetic calorimeter and a brass and scintillator sampling hadron calorimeter surround
the inner tracking volume and  provide high-resolution measurements of energy used to
reconstruct electrons, photons, and particle jets.
The calorimetry covers the region $\abs{\eta}<5.0$,
thereby providing reliable measurements of momentum imbalance in the plane transverse to the beams.
Muons are measured in gas-ionization detectors embedded in the
steel flux return yoke of the solenoid.
A trigger system selects the most interesting collisions for analysis.
A more detailed description of the CMS detector is given in Ref.~\cite{JINST}.

For this analysis, \pp collisions are selected using triggers that require the presence of at least two leptons with large transverse momentum (\pt).
Electron candidates~\cite{EGMPAS} are reconstructed by associating tracks from the inner tracker with energy clusters in the electromagnetic
calorimeter. Muon candidates~\cite{MUOART} are reconstructed by combining information from the outer muon detector with the tracks reconstructed by the inner tracker.
Additional lepton identification criteria are applied for both lepton flavors in order to reject hadronic jets that are misidentified as leptons~\cite{EGMPAS,MUOART}.
Both electrons and muons are required to be isolated from other activity in the event.
This is achieved by imposing a maximum value of 0.15 on the ratio of the
scalar sum of supplementary track \pt\ and calorimeter transverse energy deposits
within a cone of ${ \Delta R}\equiv\sqrt{\smash[b]{{ (\Delta\eta)^2+(\Delta\phi)^2}}}<0.3$ around the lepton candidate direction,
to the transverse momentum of the candidate~\cite{ref:1105.5661}.

Event selection is applied to reject events other than those from \ttbar\ production in the dilepton final state.
Events are required to have exactly two opposite-sign, isolated leptons (\eepm, \empm, or \mmpm).
The electrons (muons) are required to have $\pt >  20\GeV$ and to lie within $\abs{\eta} < 2.5$ ($2.4$).
The reconstructed lepton trajectories must be consistent with a common interaction vertex.
Events with an $\Pep\Pem$  or $\Pgmp\Pgmm$ pair with  invariant mass
in the \cPZ-boson mass ``window'' (between 76 and  106\GeV) or below
20\GeV are removed to suppress \dy\ and heavy-flavor resonance production.

The jets and the momentum imbalance in each event are reconstructed using a particle-flow
technique~\cite{CMS-PAS-PFT-10-002}. The  anti-\kt  clustering
algorithm~\cite{antikt} with a distance parameter of 0.5 is used for jet
clustering.
Corrections are applied to the energies of the reconstructed jets, based on the results of
simulations and studies using exclusive dijet and $\gamma$+jets data~\cite{JEC}.
At least two jets with  $\pt > 30\GeV$ and  $\abs{\eta} < 2.5$,
separated  by $\Delta  R  >  0.4$ from  leptons  passing the  analysis
selection, are required in each event. At least one of these jets must be
consistent with the decay of heavy-flavor hadrons (a ``\cPqb~jet''),
identified
by the combined secondary vertex \cPqb-tagging algorithm~\cite{ref:btag}.
The algorithm is based on the reconstruction of a secondary decay vertex,
and gives a \cPqb-tagging efficiency of about 70\% (depending on jet \pt\ and $\eta$)
with misidentification probabilities of approximately 1.5\% and 20\% for jets originating from light partons (\cPqu, \cPqd, and \cPqs\ quarks, and gluons) and \cPqc~quarks, respectively.
The missing transverse energy \MET\ is defined as the magnitude of
the momentum imbalance, which is the negative of the vector sum of the momenta of all
reconstructed particles in the plane transverse to the beam.
The  \MET\ in  the event is required to exceed 40\GeV in events with same-flavor leptons, to further suppress the \dy\ background.

Simulated \ttbar\ events are generated using \MCATNLO3.41~\cite{mc@nlo},
with $m_\cPqt = 172.5\GeV$,
and showered and fragmented using \HERWIG6.520~\cite{herwig6}.
Simulations with different values of $m_\cPqt$ and the factorization and renormalization scales are produced in order to evaluate the associated systematic uncertainties.

The dilepton \ttbar\ selection classifies events with $\tau$ leptons as signal only when the $\tau$ decays leptonically.
Other \ttbar\ topologies, such as the lepton+jets and all-hadronic decays, are classified as background.
The background samples of \wjets, \dy, diboson, and single-top-quark events are generated using \MADGRAPH~\cite{madgraph} or \POWHEG~\cite{powheg},
and showered and fragmented using \PYTHIA6.4.22~\cite{Pythia}.
Next-to-leading order (NLO) cross sections are used for all background samples.

For both signal and background events, multiple \pp interactions in the same or nearby bunch crossings (pileup) are simulated using \PYTHIA\ and superimposed on the hard collision.
Events are then simulated  using a \GEANTfour-based model~\cite{Geant} of the CMS  detector, and finally reconstructed and analyzed with the same software used to process collision data.

The trigger efficiency for dilepton events that pass the analysis selection criteria is determined using a tag-and-probe method as in Ref.~\cite{wzPAS2010}.
For the \eepm, \empm, and \mmpm channels this gives
\pt- and $\eta$-dependent
efficiencies of approximately 100\%, 95\%, and 90\%, respectively~\cite{ref:tprime}.
These efficiencies are used to weight the simulated events
to account for the trigger requirement.
The lepton selection efficiencies (reconstruction, identification, and isolation) are consistent between data and the simulation~\cite{wzPAS2010, PAPER-TOP-11-005}.
To account for the difference between the \cPqb-tagging efficiencies in data and the simulation~\cite{ref:btag}, data-to-simulation scale factors are applied for each jet in the simulated events.
CMS studies~\cite{toppT} have shown that the top-quark \pt distribution in data is softer than in the NLO simulation. Reweighting the top-quark \pt in the simulation to match the data
improves the modeling of the lepton and jet \pt distributions, and is applied to the \MCATNLO\ \ttbar\ sample used in this Letter.
Because of the dependence of the spin correlations on the \ttbar\ invariant mass, and thus the top-quark \pt, the \pt\ reweighting increases
the fraction of top-quark pairs with antiparallel spins in the simulation.
The simulation is used only for the unfolding, which is primarily sensitive to changes in acceptance, where
the effect of the \pt\ reweighting largely cancels in the ratio.
Still, the top-quark \pt\ spectrum modeling is one of the largest sources of uncertainty.

After all weights are applied, a total of 740 background events are expected. There are 9824 events observed in the data, and the remaining 9084 events are assumed to be signal (dileptonic \ttbar).
The average acceptance for signal events is 18\%, and describes the fraction of all produced signal events that are expected to be selected.

While the $\Delta \phi_{\ell^+\ell^-}$ measurement relies purely on leptonic information, the measurements based on $\theta^{\star}_\ell$ require the reconstruction of the entire \ttbar\  system.
Each event has two neutrinos, and there is also ambiguity in combining \cPqb~jets with leptons, resulting in up to eight possible solutions for the \ttbar\ system.
The analytical matrix weighting technique~\cite{ref:1105.5661} is used to find the most probable solution,
assuming $m_\cPqt = 172.5\GeV$. In events with only one \cPqb-tagged jet, the second \cPqb~jet is assumed to be the untagged jet with the largest \pt.
Solutions are assigned a weight, based on the probability
of observing such a configuration, and the \ttbar\ kinematic quantities are taken from the solution with the largest weight. To improve the efficiency of the technique in the presence
of mismeasured jets, the solution for each event is integrated over parametrized jet and \MET\ resolution functions.
Despite this step, ${\approx}14\%$ of the events still provide no analytic solutions,
for both the data and the simulation. These events are not used in the measurement of $\theta^{\star}_\ell$, which
is accounted for as an additional event selection requirement.

The backgrounds from \dy\ production and events with a jet misidentified as a lepton are estimated using both control data samples and simulation.
The results agree within their uncertainties.
The \dy\ background outside the \cPZ-boson mass window is estimated 
using the ratio of the number of simulated events inside the window to the number outside the window to scale
the observed event yield inside the window~\cite{ref:1105.5661}.
The contribution in this region from other processes, where the two leptons do not come from a \cPZ\ boson,
is estimated from \empm\ data and subtracted prior to performing the scaling.
The background with at least one misidentified lepton (nondileptonic \ttbar, \PW+jets, and multijet events)
is estimated
from control samples in data using a parametrization of the probability
for a jet to be misidentified as a lepton, determined using events collected with jet triggers of different energy
thresholds. For both electrons and muons, an associated ``loose" lepton candidate is defined based on relaxed isolation
requirements~\cite{ref:samesign}. The lepton misidentification rates are parametrized as a function of lepton \pt\ and $\eta$,
and are applied as weights to events containing exactly one lepton candidate and one or more loose lepton candidates.

The simulation is chosen as the method to predict the background event yields and shapes,
with systematic uncertainties
based on comparisons with the estimates using data.
The backgrounds from single-top-quark and diboson events are estimated from simulation, found in agreement with data in recent CMS measurements~\cite{ref:singletop,ref:WWWZ}.

The measured distributions are distorted from the true underlying distributions
by the limited acceptance of the detector and the finite
resolution of the measurements. An unfolding procedure is applied to correct the data for these effects, which
 yields the parton-level distributions of the variables under study,
where the full covariance matrix is used to evaluate the uncertainties and bin-to-bin correlations.

The background-subtracted measured distribution $\vec{b}$ is related to the underlying
parton-level distribution $\vec{x}$ by the matrix equation $\vec{b}=SA\vec{x}$,
where $A$ is a diagonal matrix describing the acceptance in each bin of the measured distribution and
$S$ is a smearing matrix describing the migration of events between bins due to the reconstruction techniques and finite detector resolution.
The $A$ and $S$ matrices are modeled using the \MCATNLO\ \ttbar\ simulation,
and the results are available in graphical form in \suppMaterial.

A regularized unfolding algorithm is employed using the singular value decomposition method \cite{Hocker:1995kb}.
The effects of large statistical fluctuations in the algorithm are
greatly reduced by introducing a regularization term in the unfolding procedure.
The unfolding procedure is validated using pseudoexperiments by verifying the
pull distributions and linearity for the observables under study.

Various systematic uncertainties affect the measurements. These are
mainly related to the performance of the detector and the modeling of the signal and background processes.

The uncertainty due to the jet energy scale corrections affects the analytical matrix weighting technique \ttbar\ solutions as well as the event selection.
It is estimated by varying the jet energy scale of jets within their uncertainties (typically 1\%--2\%)~\cite{JEC}, with propagation to the \MET.
The uncertainty in the lepton energy scale, which affects mainly the lepton \pt\ distributions, is estimated by varying the energy scale of electrons by 0.5\% (the uncertainty in muon energies is negligible), as estimated from comparisons between data and simulated \cPZ-boson events.

The uncertainty in the background subtraction is obtained by
varying the normalization of each background component, by 50\% for single-top-quark and diboson production and by 100\% for the backgrounds from \dy\ production and from misidentified leptons.

The \ttbar\ modeling and simulation uncertainties are evaluated by rederiving the $A$ and $S$ matrices using simulated events with variations in the parameter of interest:
the factorization and renormalization scales are together varied up and down by a factor of 2;
the top-quark mass is varied by $\pm$1\GeV\ around $m_\cPqt=172.5$\GeV;
the parton distribution functions are varied using the {\sc{pdf4lhc}} prescription~\cite{pdf4lhcInterim};
the jet energy resolution is varied by 5\%--10\%, depending on the $\eta$ of the jet~\cite{JEC};
the simulated pileup multiplicity distribution is changed within its uncertainty;
and the scale factors between data and the simulation for the \cPqb-tagging efficiency, trigger efficiency, and lepton selection efficiency are varied by their uncertainties.
In the simulated \ttbar\ events, the $\tau$ spin is not propagated correctly to its decay products. This affects the angular distributions of the electrons and muons coming from $\tau$ decays. The corresponding systematic effect is estimated by reweighting the $\tau$ decay distributions to reproduce the SM expectations.
A 100\% systematic uncertainty is applied to the top-quark \pt\ reweighting, since the origin of the effect is not yet fully understood, and the resulting systematic uncertainty is quoted separately.

Finally, the results of the unfolding linearity tests are used to estimate the systematic uncertainty in the unfolding procedure.
The contributions to the total systematic uncertainty (from their sum in quadrature) for each asymmetry variable are presented in Table~\ref{tab:asyms_sys}.

\begin{table}[!h]									
\begin{center}									
\topcaption{Systematic uncertainties in the background-subtracted and unfolded values of $A_{\Delta\phi}$, $A_{c_{1}c_{2}}$, and $A_{P}$.
}
\label{tab:asyms_sys}									
\begin{scotch}{lrrr}																		
Asymmetry variable		&	\multicolumn{1}{l}{$~~\;A_{\Delta\phi}$}	&	\multicolumn{1}{l}{$~~\;A_{c_{1}c_{2}}$}	&	\multicolumn{1}{l}{$~~\;A_{P}$}	\\	
\hline									
Jet energy scale		&$	0.002	$&$	0.012	$&$	0.009	$	\\
Lepton energy scale		&$	0.001	$&$	0.001	$&$	0.001	$	\\
Background		&$	0.003	$&$	0.001	$&$	0.006	$	\\
Fact. and renorm. scales		&$	0.001	$&$	0.010	$&$	0.004	$	\\
Top-quark mass		&$	0.001	$&$	0.003	$&$	0.005	$	\\
Parton distribution functions		&$	0.002	$&$	0.002	$&$	0.001	$	\\
Jet energy resolution		&$	{<}0.001	$&$	{<}0.001	$&$	{<}0.001	$	\\
Pileup		&$	0.002	$&$	0.002	$&$	0.004	$	\\
\cPqb-tagging scale factor		&$	{<}0.001	$&$	{<}0.001	$&$	0.001	$	\\
Lepton selection		&$	{<}0.001	$&$	{<}0.001	$&$	{<}0.001	$	\\
$\tau$ decay polarization		&$	0.001	$&$	0.002	$&$	0.001	$	\\
Unfolding		&$	0.004	$&$	0.020	$&$	0.002	$	\\
\hline									
Total systematic uncertainty		&$	0.006	$&$	0.025	$&$	0.014	$	\\
\hline	
Top \pt\ reweighting uncertainty		&$	0.012	$&$	0.010	$&$	0.008	$	\\
\end{scotch}																		
\end{center}									
\end{table}

\ifthenelse{\boolean{cms@external}}
{
\setlength{\belowcaptionskip}{-22pt}

\begin{figure}[!h]
\centering
\includegraphics[width=0.417\textwidth]{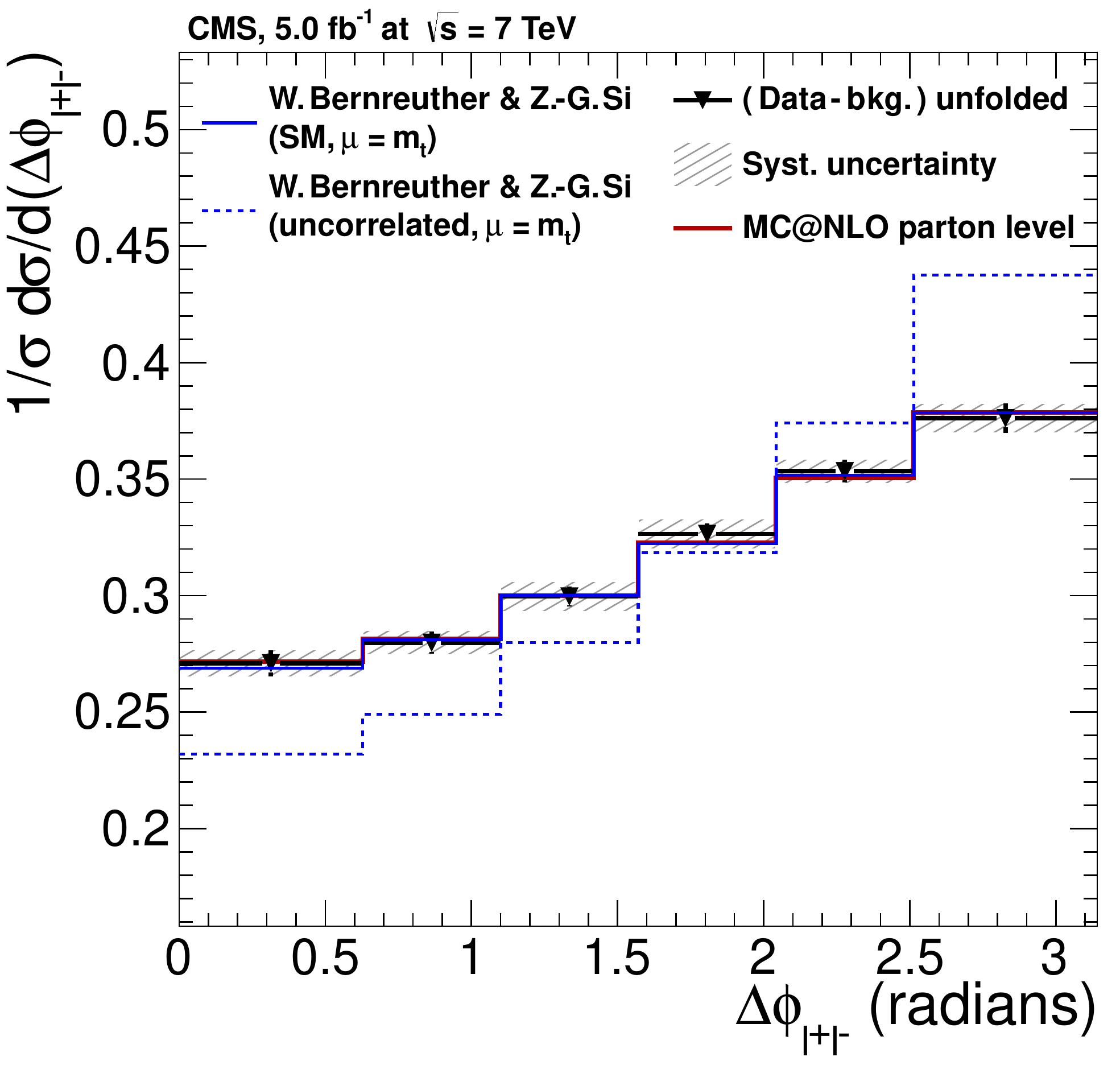}
\includegraphics[width=0.417\textwidth]{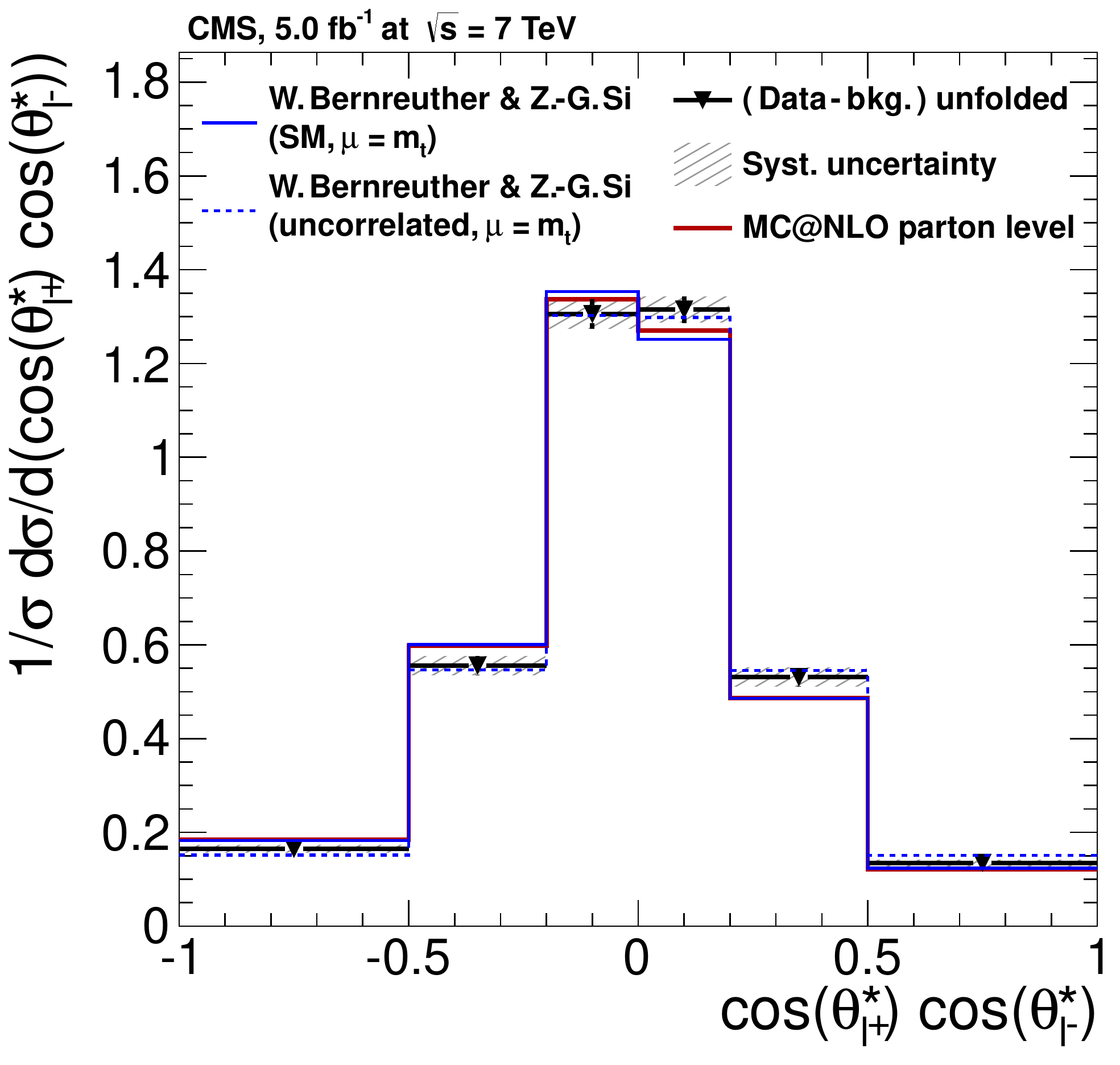}
\includegraphics[width=0.417\textwidth]{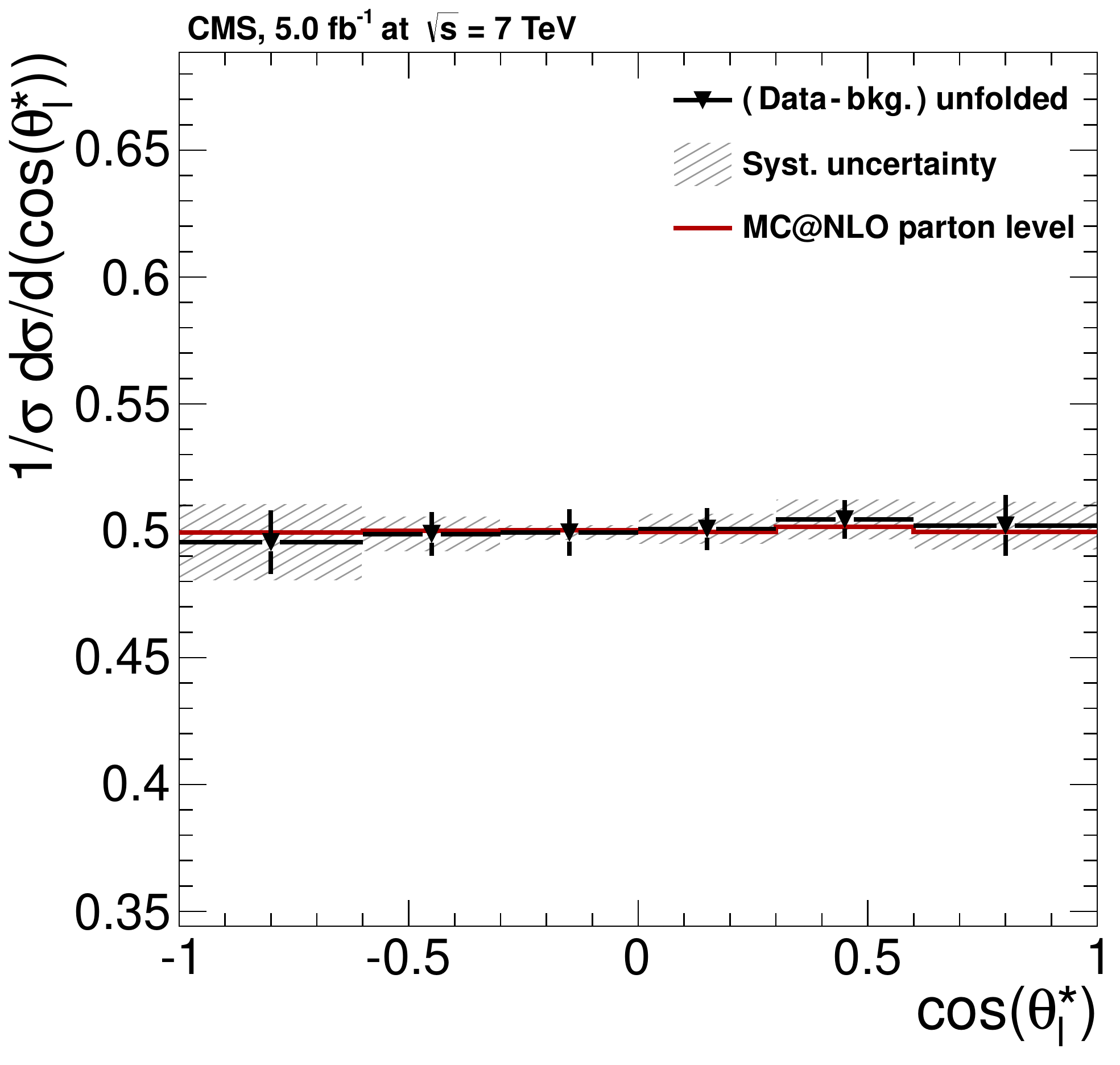}
\caption{\label{fig:ResultsUnfolded}\protect
Background-subtracted and unfolded differential cross sections for $\Delta \phi_{\ell^+\ell^-}$, $\cos(\theta^{\star}_{\ell^+}\!) \cos(\theta^{\star}_{\ell^-}\!)$, and $\cos(\theta^{\star}_\ell)$.
The error bars represent statistical uncertainties only, while the systematic uncertainty band is represented by the hatched area. The bin contents are correlated due to the unfolding.
}
\end{figure}

\setlength{\belowcaptionskip}{2pt}

}
{
\begin{figure}[!h]
\centering
\includegraphics[width=0.49\textwidth]{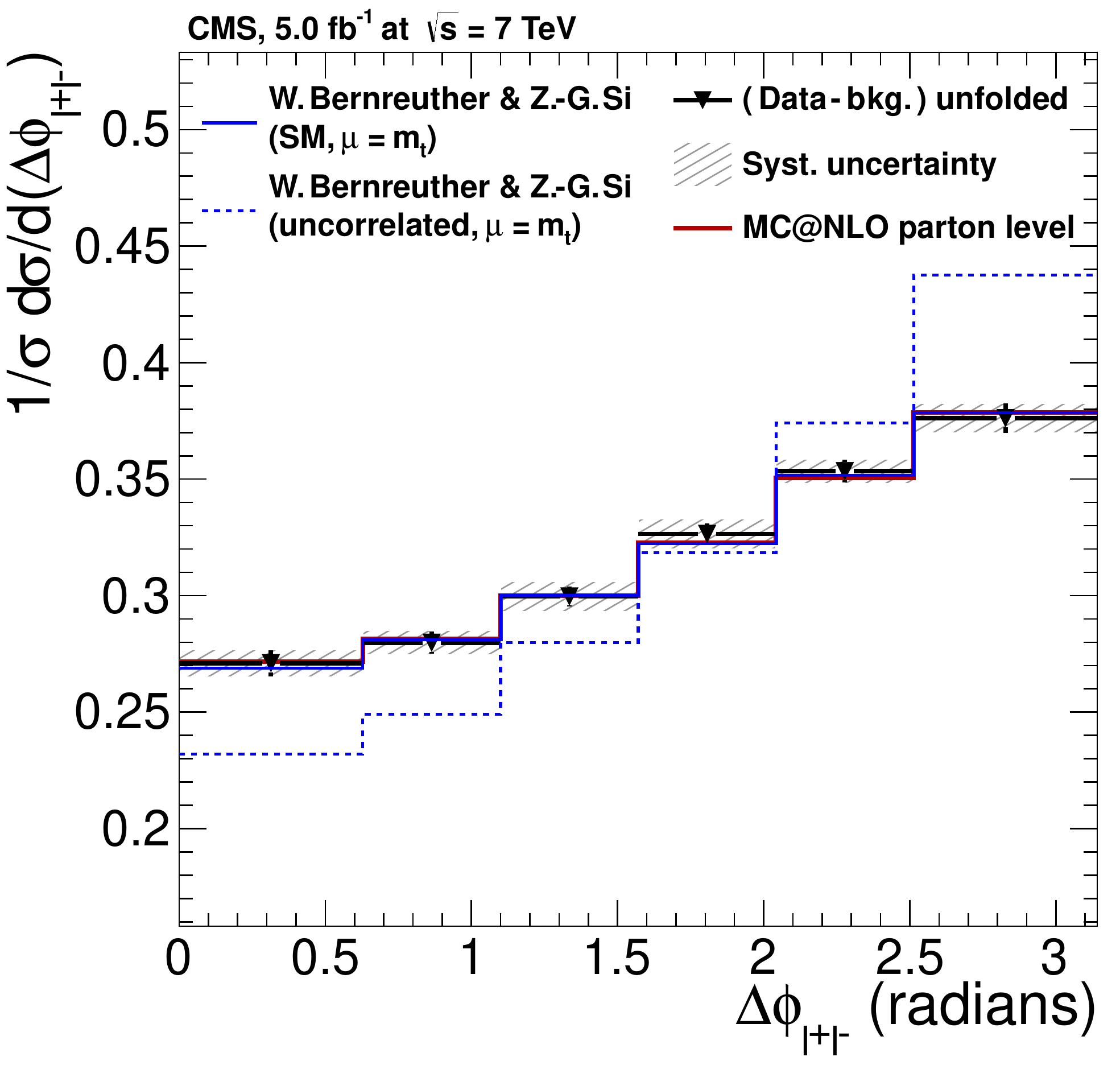}
\includegraphics[width=0.49\textwidth]{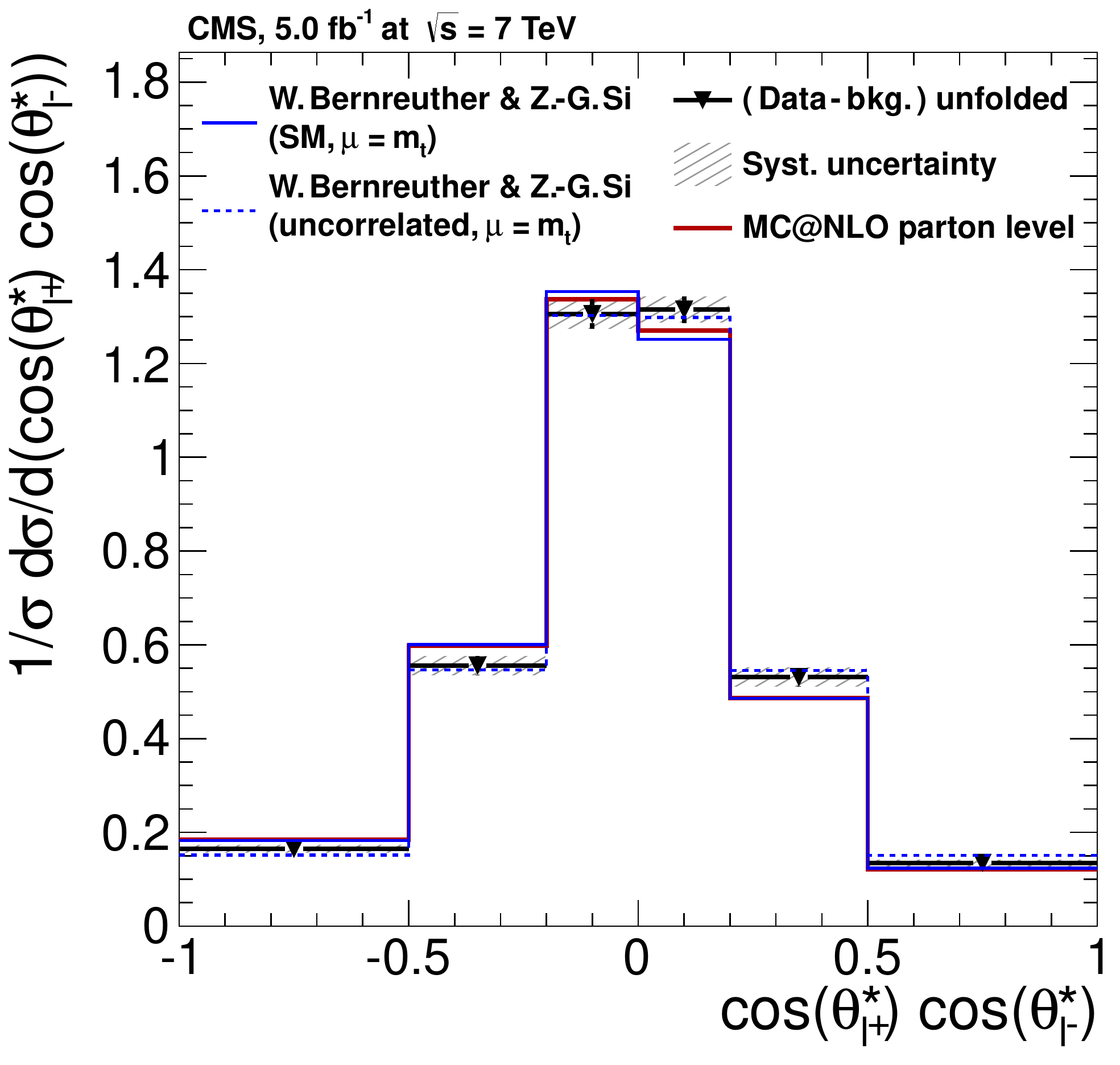}
\includegraphics[width=0.49\textwidth]{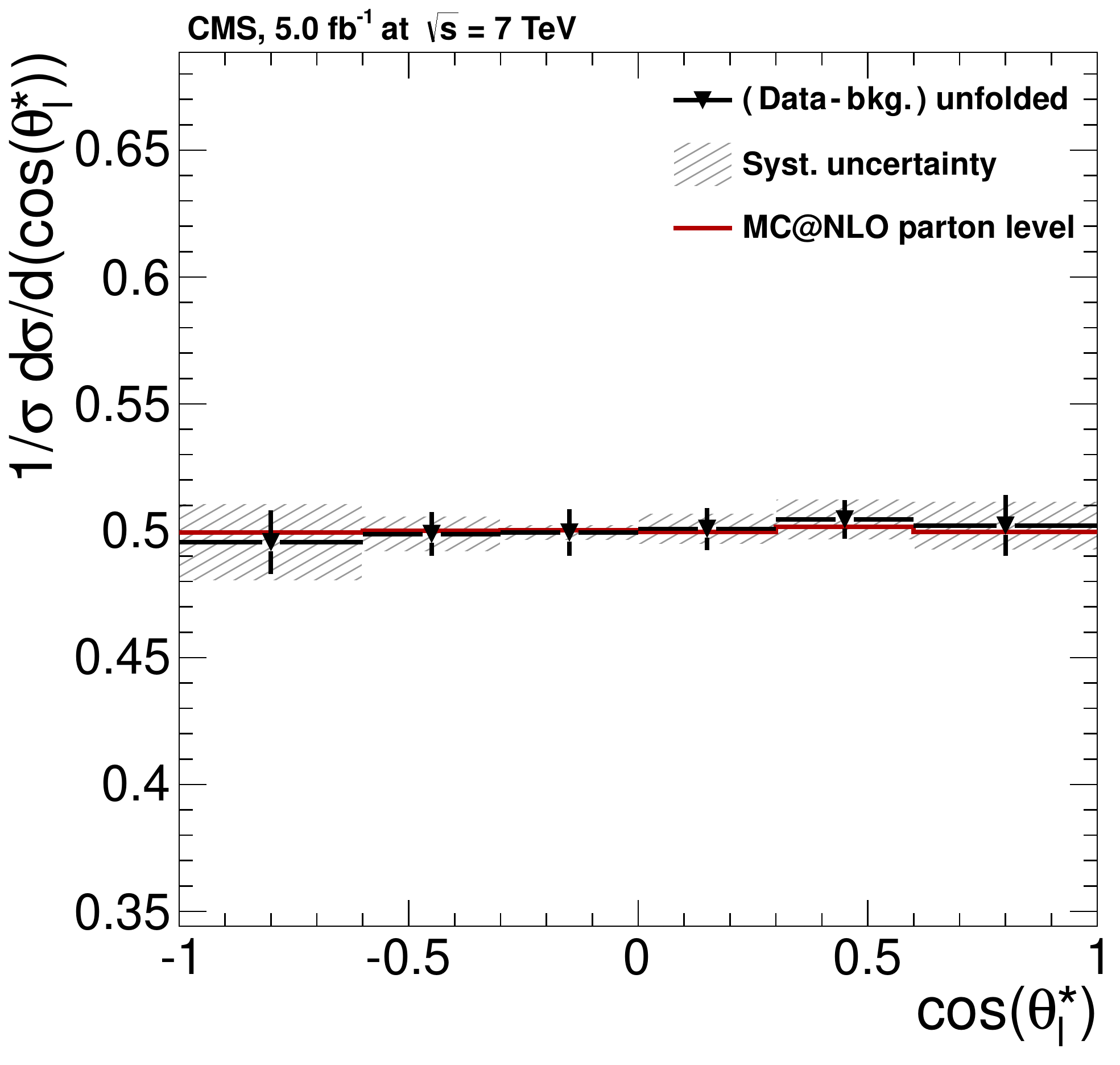}
\caption{\label{fig:ResultsUnfolded}\protect
Background-subtracted and unfolded differential cross sections for $\Delta \phi_{\ell^+\ell^-}$, $\cos(\theta^{\star}_{\ell^+}\!) \cos(\theta^{\star}_{\ell^-}\!)$, and $\cos(\theta^{\star}_\ell)$.
The error bars represent statistical uncertainties only, while the systematic uncertainty band is represented by the hatched area. The bin contents are correlated due to the unfolding.
}
\end{figure}
}

\begin{table*}[!htpb]
\centering
\topcaption{\label{tab:ResultsUnfolded}
Parton-level asymmetries. The uncertainties in the unfolded results are statistical, systematic, and the additional uncertainty from the top-quark \pt\ reweighting.
The uncertainties in the simulated results are statistical only, while the uncertainties in the NLO calculations for correlated and uncorrelated \ttbar\ spins come from scale variations up and down by a factor of two.
The prediction for $A_{c_{1}c_{2}}$ is exactly zero in the absence of spin correlations by construction.
}
\resizebox{\linewidth}{!}
{
\begin{scotch}{lrrrr}
Asymmetry  &   \multicolumn{1}{c}{Data (unfolded)} &  \multicolumn{1}{c}{\MCATNLO} & \multicolumn{1}{c}{NLO (SM, correlated)} & \multicolumn{1}{c}{NLO (uncorrelated)} \\
\hline & & & & \\ [-2.0ex]
$A_{\Delta\phi}$ &   $0.113 \pm 0.010  \pm 0.006 \pm 0.012$  &  $ 0.110 \pm 0.001$  &  $0.115^{+0.014}_{-0.016}$  &  $0.210^{+0.013}_{-0.008}$    \\ [0.5ex]
$A_{c_{1}c_{2}}$ & $-0.021 \pm 0.023 \pm 0.025 \pm 0.010$  &  $ -0.078 \pm 0.001$  &  $-0.078 \pm 0.006$ &  $0$ \\ [0.5ex]
$A_{P}$ & $0.005 \pm 0.013 \pm 0.014 \pm 0.008$  & $0.000 \pm 0.001$ & N/A & N/A \\
\end{scotch}
}
\end{table*}

The background-subtracted and unfolded distributions for $\Delta \phi_{\ell^+\ell^-}$, $\cos(\theta^{\star}_{\ell^+}\!) \cos(\theta^{\star}_{\ell^-}\!)$, and $\cos(\theta^{\star}_\ell)$ are shown in Fig.~\ref{fig:ResultsUnfolded},
normalized to unit area so that they represent parton-level differential cross sections in each variable.
The data are compared to the predictions of the \MCATNLO\ \ttbar\ sample, and
to NLO calculations for \ttbar\ production with and without spin correlations~\cite{theoretical,Bernreuther2013115}.

The asymmetries determined from the unfolded distributions are also parton-level quantities, and are measured to be
$A_{\Delta\phi} = 0.113 \pm 0.010  \pm 0.006 \pm 0.012$, $A_{c_{1}c_{2}} = -0.021 \pm 0.023 \pm 0.025 \pm 0.010$, and $A_{P} = 0.005 \pm 0.013 \pm 0.014 \pm 0.008$,
where the uncertainties are statistical, systematic, and from top-quark \pt\ reweighting, respectively.
These results are compared to the simulated and theoretical~\cite{theoretical,Bernreuther2013115} values in Table~\ref{tab:ResultsUnfolded}.
The $A_{\Delta\phi}$ result indicates the presence of \ttbar\ spin correlations, and strongly disfavors the uncorrelated case.

In summary, this Letter presents measurements related to \ttbar\ spin correlations and the top-quark polarization in the \ttbar\ dilepton final states (\eepm, \empm, and \mmpm),
using asymmetry distributions
unfolded to the parton level.
The results are in agreement with the standard model predictions for all three measured variables.

We would like to thank Professor W.~Bernreuther and Professor Z.-G.~Si for calculating the theoretical predictions of Fig.~\ref{fig:ResultsUnfolded} and Table~\ref{tab:ResultsUnfolded} for this Letter.
We congratulate our colleagues in the CERN accelerator departments for the excellent performance of the LHC and thank the technical and administrative staff at CERN and at other CMS institutes for their contributions to the success of the CMS effort. In addition, we gratefully acknowledge the computing centers and personnel of the Worldwide LHC Computing Grid for delivering so effectively the computing infrastructure essential to our analyses.  Finally, we acknowledge the enduring support for the construction and operation of the LHC and the CMS detector provided by the following funding agencies:
BMWF and FWF (Austria);
FNRS and FWO (Belgium);
CNPq, CAPES, FAPERJ, and FAPESP (Brazil);
MES (Bulgaria);
CERN;
CAS, MoST, and NSFC (China);
COLCIENCIAS (Colombia);
MSES and CSF (Croatia);
RPF (Cyprus);
MoER, SF0690030s09 and ERDF (Estonia);
Academy of Finland, MEC, and HIP (Finland);
CEA and CNRS/IN2P3 (France);
BMBF, DFG, and HGF (Germany);
GSRT (Greece);
OTKA and NIH (Hungary);
DAE and DST (India);
IPM (Iran);
SFI (Ireland);
INFN (Italy);
NRF and WCU (Republic of Korea);
LAS (Lithuania);
MOE and UM (Malaysia);
CINVESTAV, CONACYT, SEP, and UASLP-FAI (Mexico);
MBIE (New Zealand);
PAEC (Pakistan);
MSHE and NSC (Poland);
FCT (Portugal);
JINR (Dubna);
MON, RosAtom, RAS and RFBR (Russia);
MESTD (Serbia);
SEIDI and CPAN (Spain);
Swiss Funding Agencies (Switzerland);
NSC (Taipei);
ThEPCenter, IPST, STAR and NSTDA (Thailand);
TUBITAK and TAEK (Turkey);
NASU (Ukraine);
STFC (United Kingdom);
DOE and NSF (USA).

\bibliography{auto_generated}

\ifthenelse{\boolean{cms@external}}{}{
\clearpage
\appendix
\section[Supplemental material]{Supplemental material\label{app:suppMat}}

\begin{figure*}[!htpb]
\begin{center}
\includegraphics[width=0.325\linewidth]{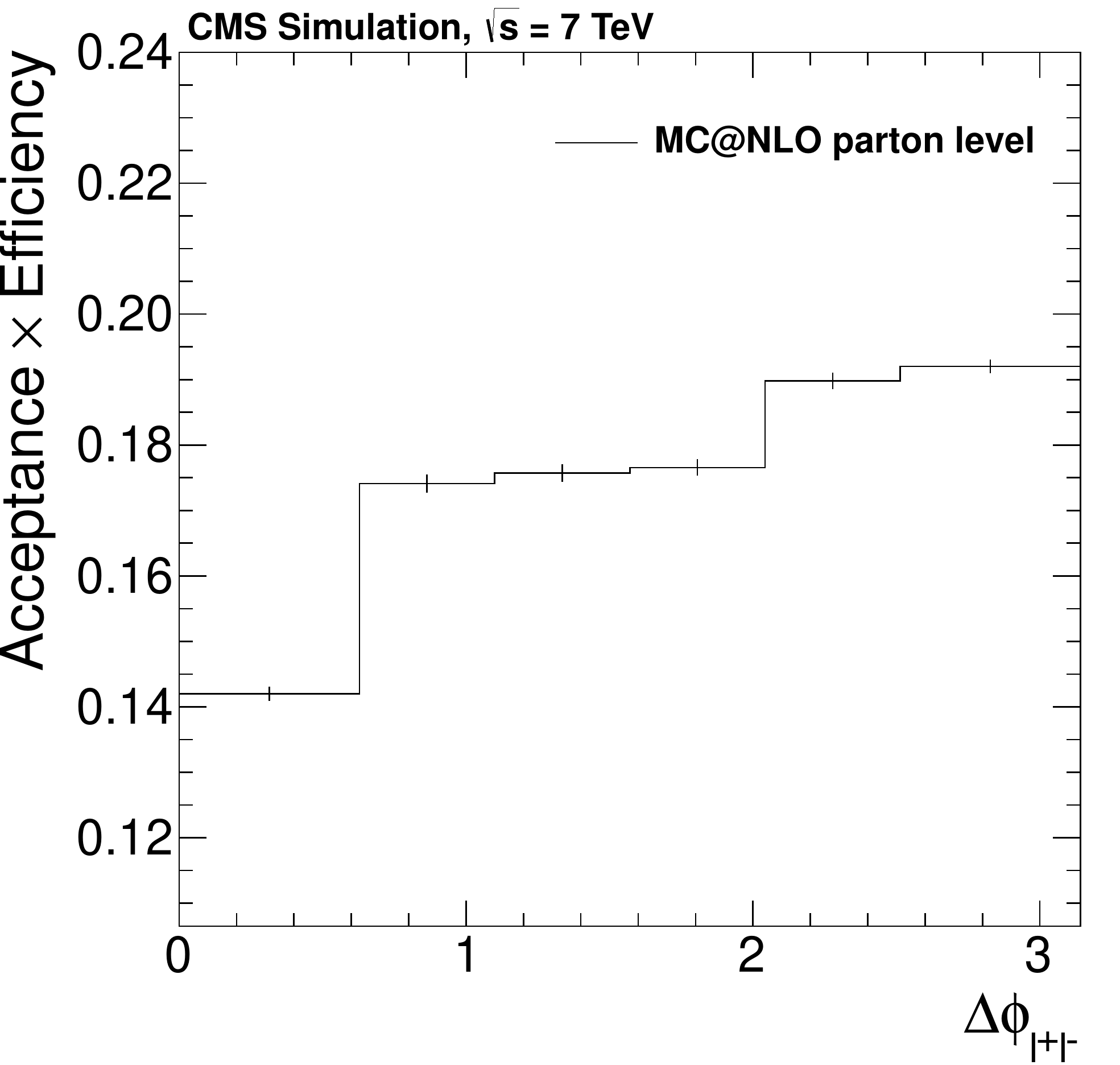}
\includegraphics[width=0.325\linewidth]{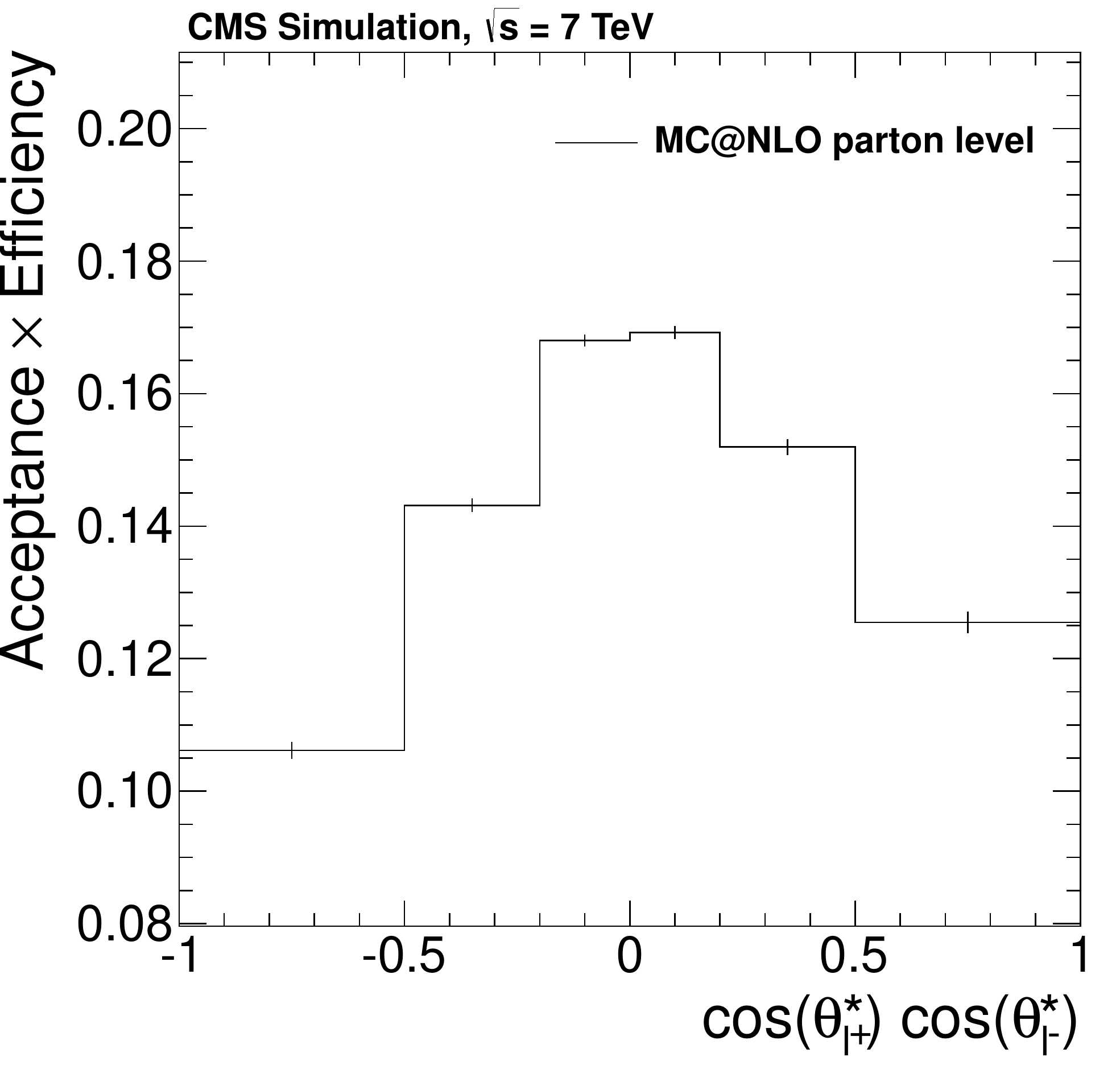}
\includegraphics[width=0.325\linewidth]{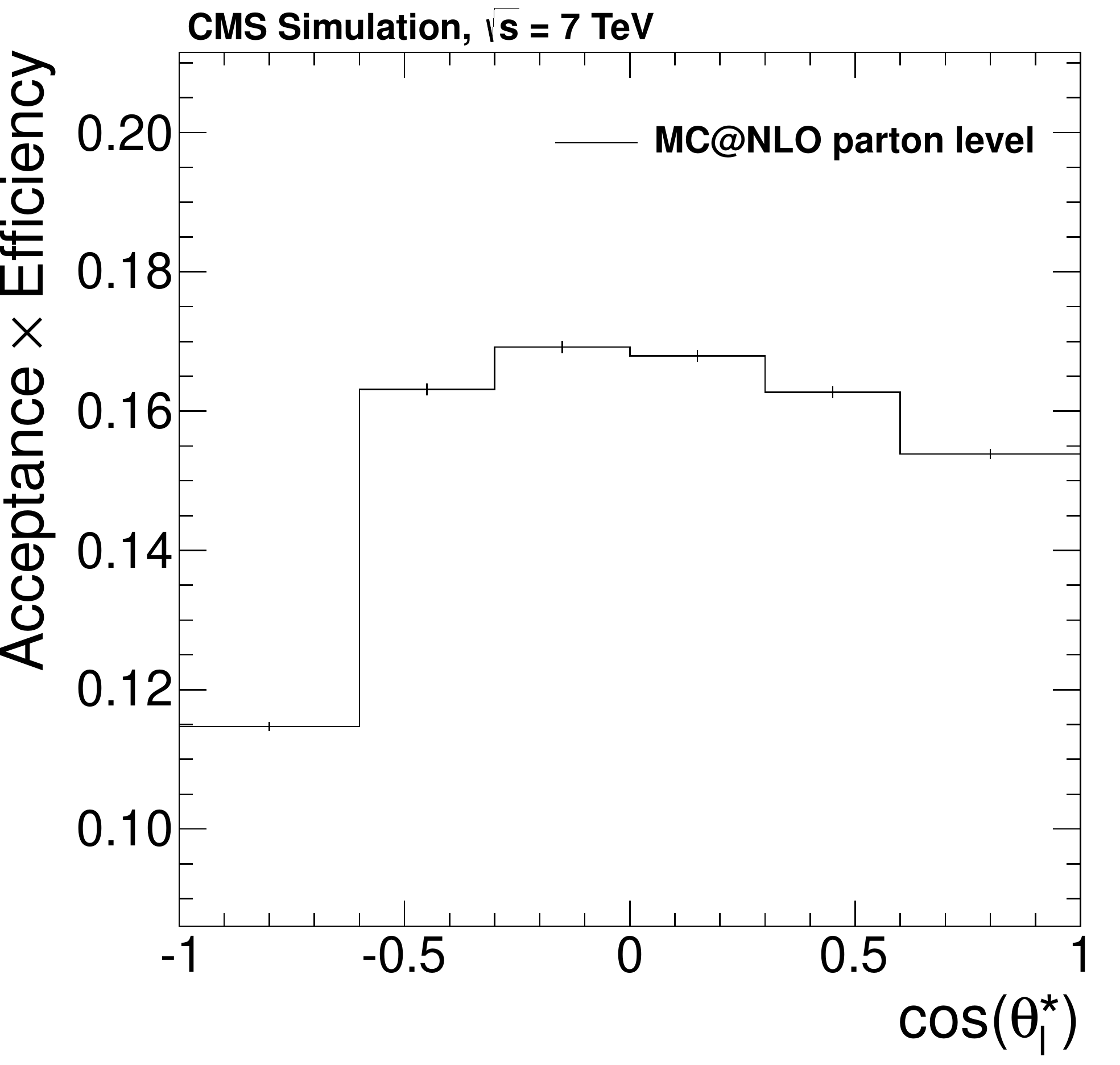}
\caption{\label{fig:AcceptMatrix}\protect
Diagonal elements of the matrix $A$ describing the acceptance times efficiency of signal events as a function of $\Delta \phi_{\ell^+\ell^-}$ (left), $\cos(\theta^{\star}_{\ell^+}\!) \cos(\theta^{\star}_{\ell^-}\!)$ (center), and $\cos(\theta^{\star}_\ell)$ (right) from simulated \MCATNLO\ \ttbar\ events. The uncertainties are statistical.
}
\end{center}
\end{figure*}

\begin{figure*}[!htpb]
\begin{center}
\includegraphics[width=0.325\linewidth]{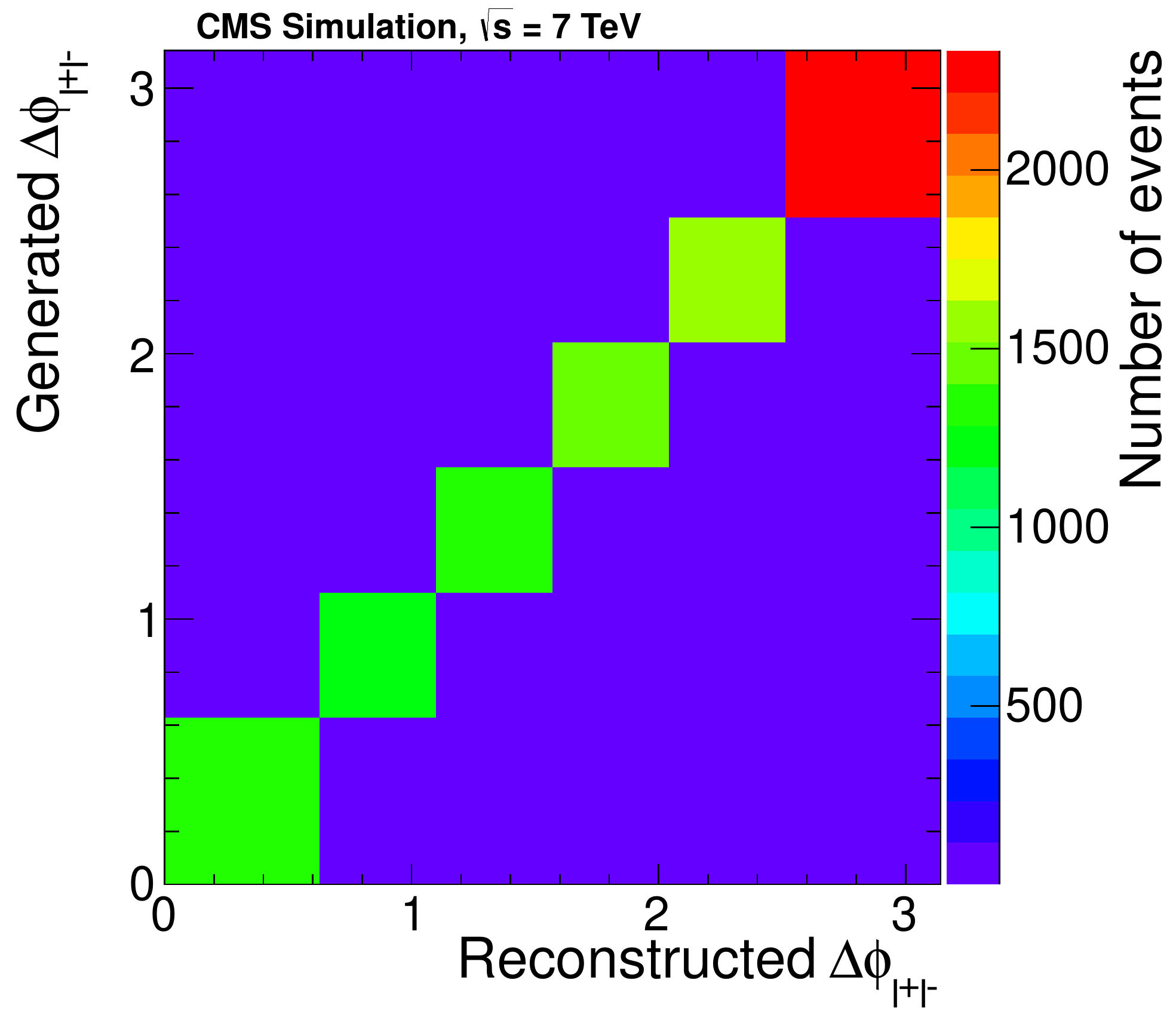}
\includegraphics[width=0.325\linewidth]{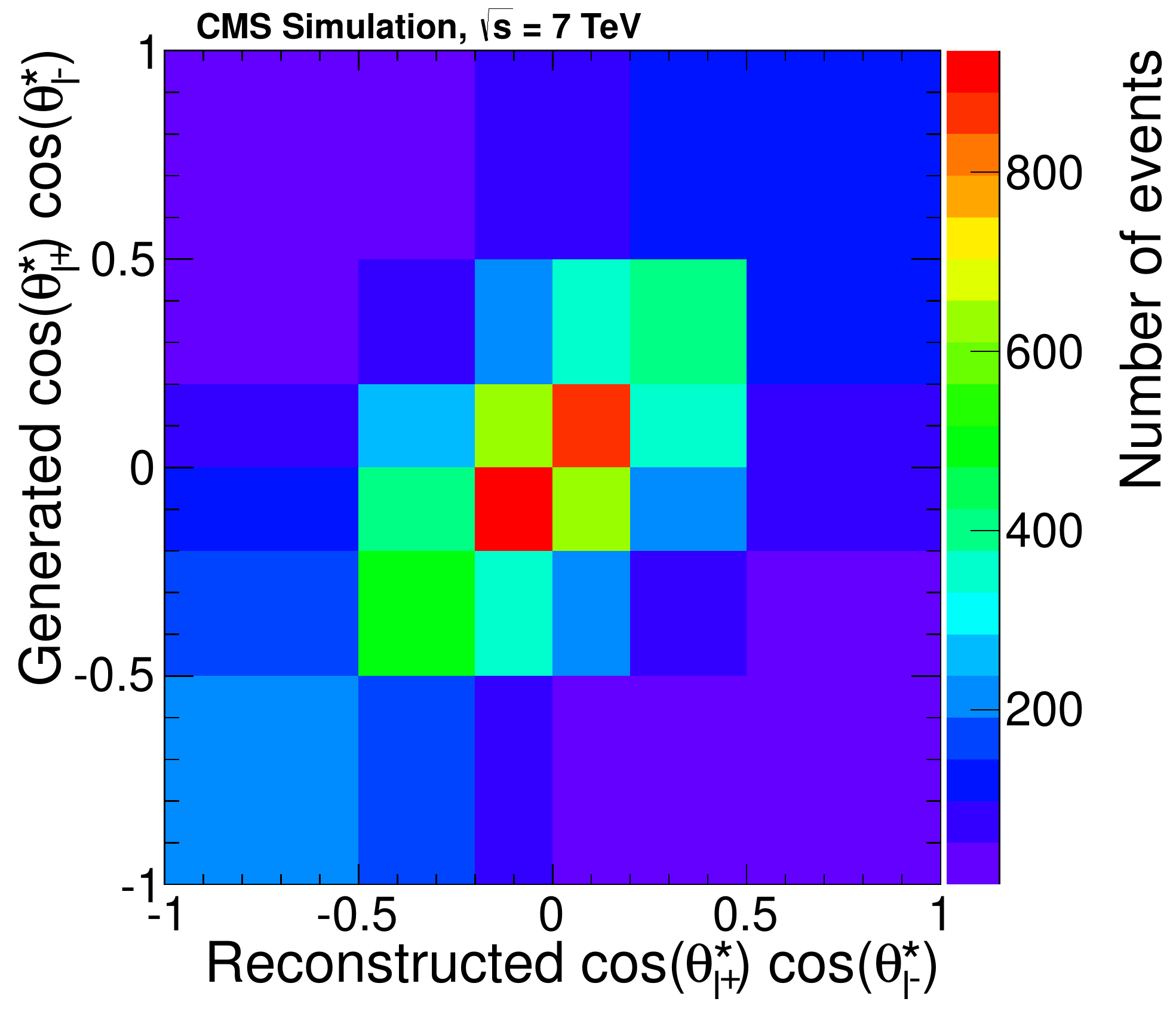}
\includegraphics[width=0.325\linewidth]{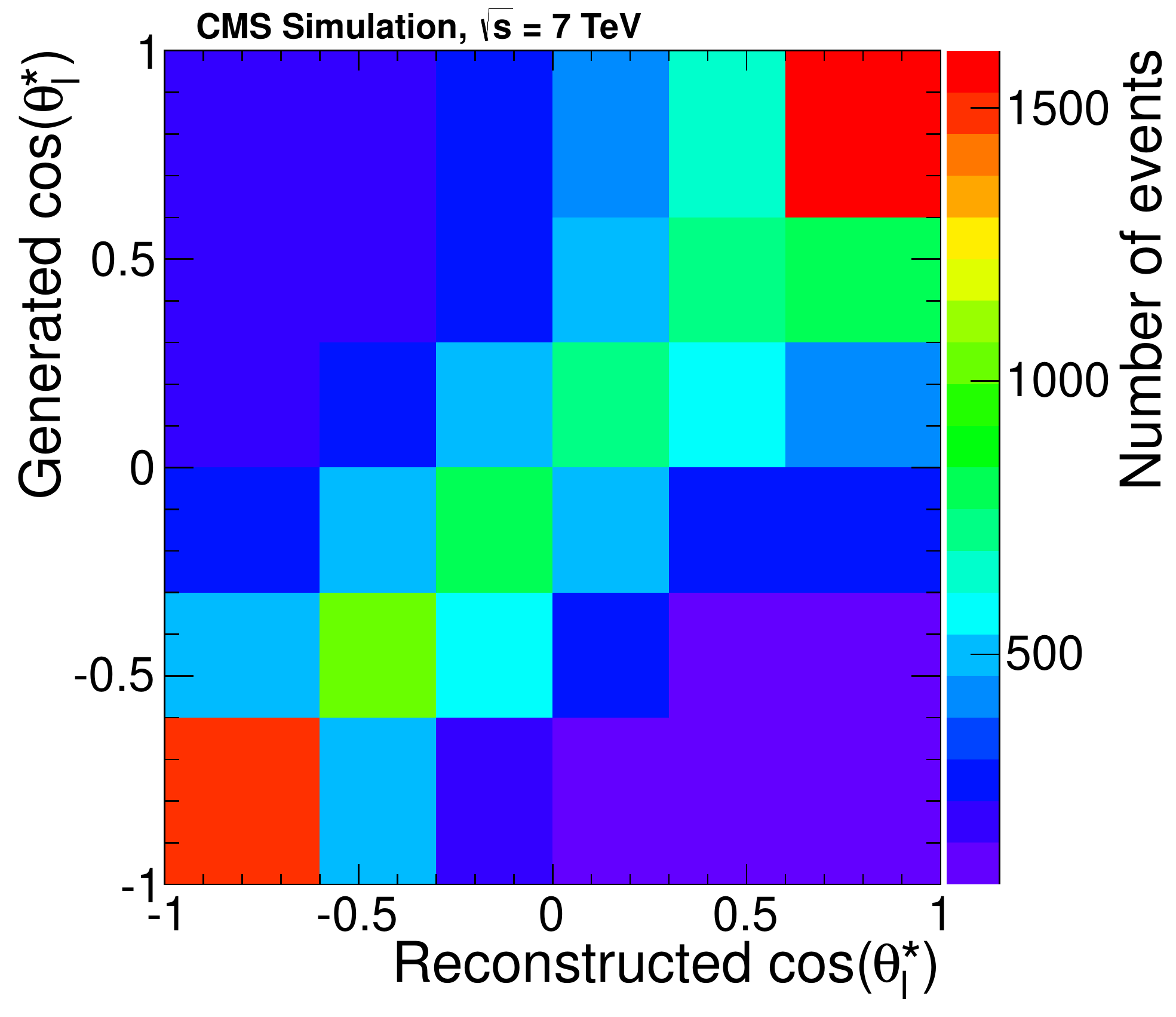}
\caption{\label{fig:SmearMatrix}\protect
Binned distributions of generated versus reconstructed values of $\Delta \phi_{\ell^+\ell^-}$ (left), $\cos(\theta^{\star}_{\ell^+}\!) \cos(\theta^{\star}_{\ell^-}\!)$ (center), and $\cos(\theta^{\star}_\ell)$ (right) from simulated \MCATNLO\ \ttbar\ events, used to derive the smearing matrices ($S$).
}
\end{center}
\end{figure*}

}

\cleardoublepage \section{The CMS Collaboration \label{app:collab}}\begin{sloppypar}\hyphenpenalty=5000\widowpenalty=500\clubpenalty=5000\textbf{Yerevan Physics Institute,  Yerevan,  Armenia}\\*[0pt]
S.~Chatrchyan, V.~Khachatryan, A.M.~Sirunyan, A.~Tumasyan
\vskip\cmsinstskip
\textbf{Institut f\"{u}r Hochenergiephysik der OeAW,  Wien,  Austria}\\*[0pt]
W.~Adam, T.~Bergauer, M.~Dragicevic, J.~Er\"{o}, C.~Fabjan\cmsAuthorMark{1}, M.~Friedl, R.~Fr\"{u}hwirth\cmsAuthorMark{1}, V.M.~Ghete, C.~Hartl, N.~H\"{o}rmann, J.~Hrubec, M.~Jeitler\cmsAuthorMark{1}, W.~Kiesenhofer, V.~Kn\"{u}nz, M.~Krammer\cmsAuthorMark{1}, I.~Kr\"{a}tschmer, D.~Liko, I.~Mikulec, D.~Rabady\cmsAuthorMark{2}, B.~Rahbaran, H.~Rohringer, R.~Sch\"{o}fbeck, J.~Strauss, A.~Taurok, W.~Treberer-Treberspurg, W.~Waltenberger, C.-E.~Wulz\cmsAuthorMark{1}
% ...
\vskip\cmsinstskip
% ...
1:~~Also at Vienna University of Technology, Vienna, Austria\\
2:~~Also at CERN, European Organization for Nuclear Research, Geneva, Switzerland\\
% ...

\end{sloppypar}
\end{document}